\documentclass[
   ,final            % use final for the camera ready runs
  ,numberedheadings % uncomment this option for numbered sections
  ]
  {aipproc}

\layoutstyle{6x9}

\usepackage{amsmath, amssymb}
\usepackage{pifont}

\begin{document}

\title{A search for wide brown dwarf companions to stars within 10pc}


\author{Marin Treselj}{
  address={Institut f\"ur Astrophysik, Georg-August-Universit\"at, D-37077 G\"ottingen, Germany}
}

\author{Andreas Seifahrt}{
 address={Institut f\"ur Astrophysik, Georg-August-Universit\"at, D-37077 G\"ottingen, Germany}
}

\author{Klaus-Werner Hodapp}{
address={Institute for Astronomy, University of Hawai{'}i, 640 N. A{'}ohoku Place, Hilo, HI 96720, USA}
}

\author{Ana Bedalov}{
address={Astrophysikalisches Institut und Universit\"ats-Sternwarte, Universit\"at Jena, Schillerg\"a\ss{}chen 2-3, 07745 Jena, Germany}
}

\author{Markus Mugrauer}{
address={Astrophysikalisches Institut und Universit\"ats-Sternwarte, Universit\"at Jena, Schillerg\"a\ss{}chen 2-3, 07745 Jena, Germany}
}

\begin{abstract}
We present the first results of a large imaging survey to identify wide brown dwarf companions
to stars within 10\,pc. We have performed a deep (\textit{H}-band limit $\sim$19.0\,mag), wide field
(up to 16$\times$16 arcmin) astrometric imaging campaign in two epochs around more than 230 nearby stars.

Preliminary results show that the wide low-mass companion fraction is far lower than
expected, indicating that interactions with the galactic disk may have removed the weakly
bound wide companions around old stars.
\end{abstract}

\maketitle

%%%%%%%%%%%%%%%%%%%%%%%%%%%%%%%%%%%%%%%%%%%%
%% MAINMATTER
%%%%%%%%%%%%%%%%%%%%%%%%%%%%%%%%%%%%%%%%%%%%

\section{The sample}
At the beginning of the campaign in 2004, we have
carried out a literature study to collect a complete
sample of all known stars within 10\,pc. We found
274 targets with final or tentative parallaxes larger
than 100\,mas. 232 of these stars were successfully
observed from two sites:

128 stars on the northern and southern hemisphere
were selected for observations with the ULBCAM at
the UH88 telescope on Mauna Kea, Hawai{'}i. Two
epochs were obtained for all stars in multiple runs
between September 2005 and February 2008.
Depending on the seeing conditions, an average
limiting magnitude of $\sim$19\,mag was achieved in \textit{H}-band.
The FOV of ULBCAM is about 16$\times$16 arcmin.

104 stars with low southern declinations were
imaged in two epochs with SOFI at the 3.5m NTT on
La Silla, Chile. An average depth of 
H$_\mathrm{limit}\sim19$\,mag was achieved in a 
field of 5$\times$5 arcmin. Observations
were carried out in four runs between January 2005
and June 2006.
Both surveys are sensitive to brown dwarfs with
masses as low as $\sim$25\,M$_\mathrm{Jup}$ at an 
age of 5\,Gyrs (according to theoretical isochrones 
by Baraffe et al. \cite{2003A&A...402..701B}).
This survey extends previous campaigns (e.g., \cite{2002AJ....123.2027H}) 
and complements coronographic search programs for close companions
(e.g., \cite{2001AJ....121.2189O}).

\section{Data reduction}

Images with the ULBCAM were taken in read-resetread
mode. Data reduction followed the standard
technique of flatfielding and background subtraction.
The large FOV of the camera leads to
noticeable field distortions which had to be
corrected before co-adding the individual frames.
We used the \textsl{TERAPIX} \cite{2002ASPC..281..228B} packages \textsl{SExtractor}, \textsl{Scamp}
and \textsl{Swarp} to correct the distortions (see Fig.~1) , to calibrate
the astrometry of the images (see Fig.~2), and finally combining
them to a full mosaic.

\begin{figure}[ht]
\resizebox{\textwidth}{!}{
\includegraphics[angle=270,bb=40 0 532 780, clip]{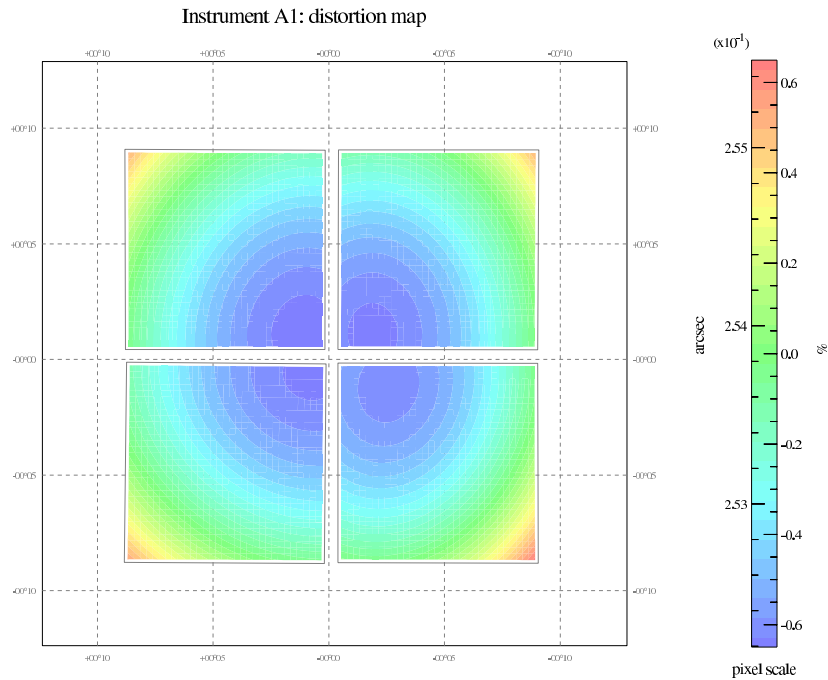}
}
\caption{Distortion map of the four chips of the ULBCAM. Correcting
the distortions is necessary to combine all images from a large
dither pattern and to provide a robust astrometry over the full
FOV of the camera.}
\end{figure}

\section{Preliminary results}

Two final images of the same field were obtained for
each of the 232 targets with an epoch difference of one
year. Object detections were performed with \textsl{SExtractor}
and \textsl{IDL/STARFINDER}. The two source catalogues were
than compared to identify objects with a proper motion
identical to the nearby star in the same field. In addition,
visual inspection of both images (the classical ``blinking''
method) was used to ensure results in crowded fields.

The ongoing analysis of the complete southern sample
(100 fields obtained with SOFI at the 3.5m NTT) and the
first 20 images of the northern sample revealed no new
companion candidate. This seems to indicate a much
lower fraction of low-mass wide companions among the
stars in the solar neighbourhood than found in younger
clusters and star forming regions. 

A possible scenario to explain the deficiency of weakly 
bound companions is the enhanced gravitational interaction 
of old stars with the galactic disk. Hence, the expected 
low mass companions of these stars were teared away and 
are no longer co-moving with their stellar counterpart.

\begin{figure}[ht]
\resizebox{.6\textwidth}{!}
{\includegraphics[bb=40 150 534 599, angle=270,clip]{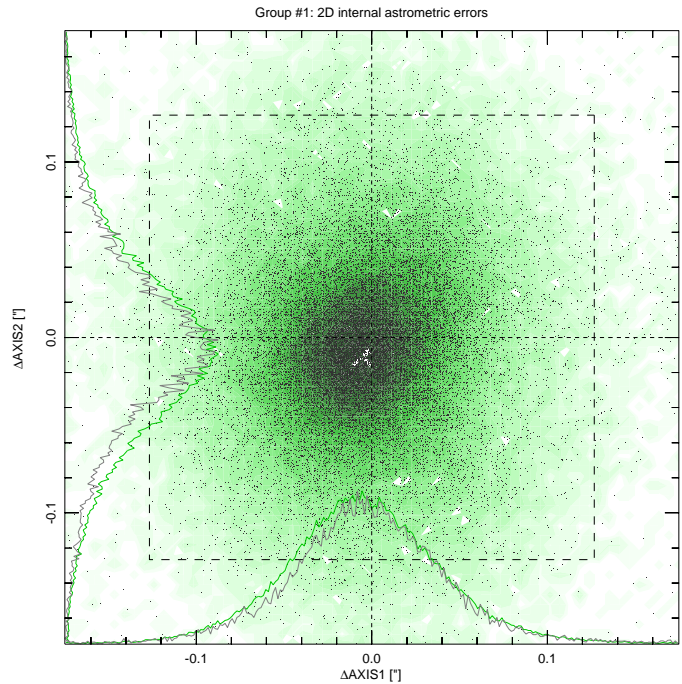}}
  \caption{Pattern matching of source detections over the full FOV
from all individual frames of one target field. The dashed box
marks the size of one ULBCAM pixel (250\,mas). The FWHM of the
distribution stays well below 100\,mas and demonstrates the
astrometric quality of the final image.}
\end{figure}

%\begin{figure}
%\resizebox{\textwidth}{!}
%{\includegraphics[bb = -48 120 661 713,clip]{GJ860_poster}}
%  \caption{An example of the data obtained with the ULBCAM, the rich
%stellar field around GJ 860. GJ860 is partially hidden behind the
%gap of the detector mosaic. The total FOV is about 23$\times$23\,
%arcmin. The full depth of H$_\mathrm{limit}\sim19$\,mag is reached 
%in the central 16$\times$16\,arcmin.}
%\end{figure}

%%%%%%%%%%%%%%%%%%%%%%%%%%%%%%%%%%%%%%%%%%%%%%%%
%% BACKMATTER
%%%%%%%%%%%%%%%%%%%%%%%%%%%%%%%%%%%%%%%%%%%%%%%%

%\begin{theacknowledgments}
%The conference organisors.
%\end{theacknowledgments}

%%%%%%%%%%%%%%%%%%%%%%%%%%%%%%%%%%%%%%%%%%%%%%%%
%% The bibliography can be prepared using the BibTeX program or
%% manually.
%%
%% The code below assumes that BibTeX is used.  If the bibliography is
%% produced without BibTeX comment out the following lines and see the
%% aipguide.pdf for further information.
%%
%% For your convenience a manually coded example is appended
%% after the \end{document}
%%%%%%%%%%%%%%%%%%%%%%%%%%%%%%%%%%%%%%%%%%%%%%%%

%%%%%%%%%%%%%%%%%%%%%%%%%%%%%%%%%%%%%%%%%%%%%%%%
%% You may have to change the BibTeX style below, depending on your
%% setup or preferences.
%%
%%
%% For The AIP proceedings layouts use either
%%%%%%%%%%%%%%%%%%%%%%%%%%%%%%%%%%%%%%%%%%%%

\bibliographystyle{aipproc}   % if natbib is available
%\bibliographystyle{aipprocl} % if natbib is missing

%%%%%%%%%%%%%%%%%%%%%%%%%%%%%%%%%%%%%%%%%%%
%% You probably want to use your own bibtex database here
%%%%%%%%%%%%%%%%%%%%%%%%%%%%%%%%%%%%%%%%%%%
%\bibliography{sample}

%%%%%%%%%%%%%%%%%%%%%%%%%%%%%%%%%%%%%%%%%%%
%% Just a reminder that you may have to run bibtex
%% All of it up to \end{document} can be removed
%% if you don't like the warning.
%%%%%%%%%%%%%%%%%%%%%%%%%%%%%%%%%%%%%%%%%%%
\IfFileExists{\jobname.bbl}{}
 {\typeout{} \typeout{******************************************}
  \typeout{** Please run "bibtex \jobname" to optain} \typeout{** the
  bibliography and then re-run LaTeX} \typeout{** twice to fix the
  references!}  \typeout{******************************************}
  \typeout{} }

\end{document}